\documentclass[%
 reprint,
 amsmath,amssymb,
 aps,
 natbib=false
]{revtex4-2}
\usepackage{graphicx}
\usepackage{dcolumn}
\usepackage{bm}
\usepackage[utf8]{inputenc} 
\usepackage[T1]{fontenc}    
\usepackage{textalpha}
\usepackage[ngerman,english]{babel}
\usepackage[titletoc,title]{appendix}

\usepackage{amsfonts}  
\usepackage{amsmath}   
\usepackage{amssymb}   
\usepackage{fancyhdr}
\usepackage{parskip}
\usepackage{subcaption}
\usepackage{multirow}
\usepackage{mathtools}
\usepackage{leftindex}
\usepackage{xcolor}

\newcommand{\pwisein}{\left\{ \begin{array}{ll}}
\newcommand{\pwiseout}{\end{array}\right.}

\def\thuphys{Department of Physics and Center for High Energy Physics, Tsinghua University, Beijing 100084, China}

\begin{document}

\preprint{APS/123-QED}

\title{Extended Variable Phase Method for Spin-1/2 Correlation Functions}

\author{Renjie Zou}
 \email{zourj24@mails.tsinghua.edu.cn}
\affiliation\thuphys
\author{Sheng Xiao}
\affiliation\thuphys
\author{Zhi Qin}
\affiliation\thuphys
\author{Zhigang Xiao}
\affiliation\thuphys


\date{\today}

\begin{abstract}
We have developed a systematic approach to calculate the correlation function for spin-1/2 particles, incorporating both central and noncentral components of the interparticle interaction. This is achieved by extending the variable phase method to accommodate noncentral potentials and numerically solving the Schrödinger equation. Within this framework, the partial-wave contributions to the nucleon-nucleon correlation functions adopting the Reid soft-core potential are evaluated. The resulting correlation functions are then compared for Gaussian sources of different sizes.
\end{abstract}

\maketitle


\section{\label{sec:level1}Introduction}
The precursor of what is now known as the correlation function, the HBT effect, was first derived by Hanbury Brown and Twiss~\cite{HBT1} in the early 1950s within the field of astronomy to estimate the angular diameters of radio stars. The underlying experimental principle is based on the intensity interference phenomenon of two electromagnetic waves emitting from an incoherent source~\cite{HBT2}. Later, this concept was extended beyond photons. In 1959, G. Goldhaber and colleagues~\cite{FirstCF} observed angular correlations of pions produced in proton-antiproton collisions, marking a milestone in the correlation function as the first interference measurements involving particles other than photons. 

Since then, the study of two-particle correlations has evolved into a dynamic and vital field. Far beyond the original purpose of extracting source sizes and interaction parameters, it now encompasses a wide range of research topics, including collision dynamics~\cite{CollisionDynamics1,CollisionDynamics2,CollisionDynamics3}, source imaging~\cite{SourceImaging1,SourceImaging2,SourceImaging3,SourceImaging4}, resonance states~\cite{ResonanceStates1,ResonanceStates2}, collective flow~\cite{CollectiveFlow1,CollectiveFlow2,CollectiveFlow3,CollectiveFlow4}, and neutron skins~\cite{SourceImaging2,NeutronSkin1}. These investigations are supported by a powerful blend of traditional theoretical frameworks and modern analysis tools, such as the Richard-Lucy algorithm~\cite{RL_algorithm}. Furthermore, the scope of correlated pairs has been extended from pions, kaons, and nucleons~\cite{OrdinaryParticles1} to include more exotic particles~\cite{ExoticParticles1,ExoticParticles2,ExoticParticles3,ExoticParticles4,ExoticParticles5,ExoticParticles6,ExoticParticles8,ExoticParticles9}, light nuclei~\cite{LightNuclei2,LightNuclei3,LightNuclei5}, and even gluon jets~\cite{Jets}.

The partial wave method~\cite{partialwavemethod} is a cornerstone of using correlation functions to constrain interaction parameters, stemming from the nature of intensity interference as a low-energy scattering experiment. For any source of finite density, the pair of particles under investigation approaches decoupling as the relative momentum tends to infinity, thereby suppressing the contribution of partial waves with large angular momentum. Consequently, the formalism of correlation functions can be significantly simplified by considering only the lowest one or few angular momenta. For the majority of previous experiments, satisfactory fitting results have been achieved using theoretical curves based solely on s-wave interactions, such as the Lednicky–Lyuboshitz model~\cite{LL1}. Recent advances in detector technology and data processing methods have gradually enabled the observation of some higher partial wave contributions. Additionally, the introduction of underlying resonance states in pair correlations has placed new demands on our ability to interpret enhancements observed in specific partial waves~\cite{ResonanceState1,ResonanceStates2,ResonanceState3}.\vspace{-2pt}

The noncentral potential, specifically the tensor force, naturally enters the formalism when considering higher partial waves in pairs with spin–orbit coupling~\cite{noncentral_basic1,noncentral_basic3,noncentral_basic2}, though its influence can sometimes also be observable at low energies. One of the most prominent examples is the existence of an electric quadrupole moment in the ground state of the deuteron, which arises from tensor-force mixing between the $^3S_1$ and $^3D_1$ channels. In the Reid soft-core potential for nucleons~\cite{ReidPotential}, the lowest orbital angular momentum at which the tensor potential appears is the s-wave for total isospin $T=0$ and the p-wave for $T=1$, with comparable magnitudes, which implies a non-negligible contribution to some correlation functions. Despite its importance in nuclear and particle physics, the tensor force has rarely been addressed in previous correlation function studies~\cite{noncentral_research}, partly due to the considerable computational cost of using shooting algorithms~\cite{ShootingMethods1,ShootingMethods2}.\vspace{0pt}

The main objective of this work is to elucidate the role of higher partial waves and noncentral potentials in shaping the fine structure of two-particle correlation functions. To this end, we introduce a novel approach based on the variable phase method. This method is capable of handling arbitrary partial waves for central interactions involving particles of arbitrary spin, as well as tensor forces for spin-1/2 particles. This formalism is successfully applied to compute nucleon-nucleon correlation functions, enabling a systematic investigation of several key topics, including the contribution of individual partial waves, the influence of source size, and the existence of recoil.\vspace{0pt}
\vspace{-5pt}\section{Methods}
\label{sec:methods}
In theoretical frameworks~\cite{LL2,AHP,KooninPratt1,KooninPratt2}, the correlation function, $C(\mathbf{q},\mathbf{K})$ for a pair of particles with relative momentum $\mathbf{q}$ and total momentum $\mathbf{K}$ is given by
\begin{eqnarray}
C(\mathbf{q},\mathbf{K})=\int d^3r \left| \Psi(\mathbf{q},\mathbf{r^*}) \right|^2S(\mathbf{r})
\label{CF}
\end{eqnarray}
where the source emission function $S(\mathbf{r})$ describes the probability of emitting a pair at a relative distance $r$ in the source rest frame, and $\Psi$ is the relative wave function of the pair after taking into account the final-state interaction. The wave function is obtained by solving the scattering problem with appropriate boundary conditions, followed by a time-reversal operation and a Lorentz transformation to the source rest frame~\cite{AHP}. The diversity of theoretical approaches to calculating correlation functions primarily stems from different methods used to derive the wave function $\Psi$.
\subsection{Wave functions in central potentials}
\label{sub:methods_cen}

For completeness, Let's recall the wave functions in central potential. The approach for obtaining accurate wave functions for a particle with reduced mass $\mu$ and incident momentum magnitude $k$ under arbitrary central potentials is based on the variable phase method, originally developed by Morse and Allis in the 1930s \cite{VPE1,VPE2,VPE3,VPE4} to calculate phase shifts in scattering processes. The derivation begins with the integral form of the radial Schrödinger equation
\begin{eqnarray}
&&u_l(r)=C\hat{j}_l(kr)+k^{-1}\int_{0}^{r}ds\left[\hat{j}_l(kr)\hat{n}_l(ks)\right.\nonumber\\
        &&\left.-\hat{j}_l(ks)\hat{n}_l(kr)\right]V(s)u_l(s),
\label{int_of_Schro}
\end{eqnarray}
where $\hat{j}_l$ and $\hat{n}_l$ denote Riccati-Bessel functions \cite{Math1,Math2}, $V(r)$ is the potential rescaled by $2\mu$, and $C$ is a constant of integration. This formula can be comprehended as a radial version of the Lippmann-Schwinger equation.
To simplify the integral form, we introduce auxiliary functions
\begin{eqnarray}
\begin{aligned}
&c_l(r)=C+k^{-1}\int_{0}^{r}dr'V(r')\hat{n}_l(kr')u_l(r')\\
&s_l(r)=-k^{-1}\int_{0}^{r}dr'V(r')\hat{j}_l(kr')u_l(r'),
\end{aligned}
\end{eqnarray}
so the radial wave function $u_l(r)$ can be substituted by
\begin{eqnarray}
u_l(r)=\hat{j}_l(kr)c_l(kr)+\hat{n}_l(kr)s_l(kr).
\label{u_l(r)}
\end{eqnarray}
Additionally, derivatives of $s_l(r)$ and $c_l(r)$ are simplified as
\begin{eqnarray}
\begin{aligned}
&s^\prime_l(r)=-k^{-1}V(r)\hat{j}_l(kr)[c_l(r)\hat{j}_l(kr)+s_l(r)\hat{n}_l(kr)],\quad\\
&c^\prime_l(r)=k^{-1}V(r)\hat{n}_l(kr)[c_l(r)\hat{j}_l(kr)+s_l(r)\hat{n}_l(kr)].\quad
\end{aligned}
\label{deriv_sc}
\end{eqnarray}
Defining $t_l(r)=s_l(r)/c_l(r)$ and differentiating yields the key differential equation in the variable phase method
\begin{eqnarray}\label{eqb}
    t_l'(r)=-k^{-1}V(r)\left[\hat{j}_l(kr)+t_l(r)\hat{n}_l(kr)\right]^2.
\end{eqnarray}
The asymptotic limit $r\to\infty$ of Eq. (\ref{u_l(r)}) relates $t_l(r)$ to the phase shift $\delta_l(\infty)$:
\begin{eqnarray}\label{eqc}
    \displaystyle\lim_{r\to\infty}t_l(r)=\tan{\delta_l(\infty)}.
\end{eqnarray}
Numerically solving Eq.~(\ref{eqb}) and (\ref{eqc}) with the boundary condition $\delta_l(0)=0$ constitutes the computational core of the variable phase method~\cite{VPE1,VPE2,VPE3,VPE4}. The term "variable phase" itself originates from extending the concept of the phase shift to finite distances by introducing the $r$-dependent function $\delta_l(r)$ via the relation $t_l(r)=\tan{\delta_l(r)}$. 

Next we define $A_l(r)$ through
\begin{eqnarray}\label{equsc}
    \begin{aligned}
    &c_l(r)=A_l(r)\cos{\delta_l}\\
    &s_l(r)=A_l(r)\sin{\delta_l},
    \end{aligned}
\end{eqnarray}
where $\delta_l=\delta_l(r)$. The radial wave function $u_l(r)$ can then be expressed as
\begin{eqnarray}
    u_l(r)=A_l(r)[\sin{\delta_l}\hat{n}_l(kr)+\cos{\delta_l}\hat{j}_l(kr)].
\label{radialwavefunc}
\end{eqnarray}
From the definition in Eq.~\eqref{equsc}, it follows that $c_l^2(r)+s_l^2(r)\equiv A_l^2(r)$. Differentiating this identity, dividing by $A_l^2(r)$, and substituting from Eq.~(\ref{equsc}) yields
\begin{eqnarray}
&&(\ln{A_l(r)})'=k^{-1}V(r)\left[ \hat{j}_l(r)\hat{n}_l(r)(\cos^2{\delta_l}-\sin^2{\delta_l})\right.\nonumber\\
&&\left.-\sin{\delta_l}\cos{\delta_l}\left(\hat{j}^2_l(r)-\hat{n}^2_l(r)\right) \right].
\label{Alr}
\end{eqnarray}
In most scattering problems, incoming spherical waves at infinity are discarded \cite{QM1,QM2}. Consequently, the specific combination coefficient is fixed as $A_l(\infty)=(2l+1)i^l\exp{[i\delta_l(\infty)]}$. Imposing this boundary condition and integrating the differential equation \eqref{Alr} governing its radial evolution yields the explicit form of $A_l(r)$, from which the radial wave function $u_l(r)$ can be directly obtained via Eq.~\eqref{radialwavefunc}.

According to the partial wave method~\cite{partialwavemethod}, the full wave function is expressed as
\begin{eqnarray}
\psi(\mathbf{r})=\displaystyle\sum_{l=0}^\infty \frac{u_l(r)}{kr}P_l(\cos\theta).
\end{eqnarray}
For practical numerical implementation, we truncate the contribution of the short range interaction in the infinite sum at angular momentum $l_0$ and recover the plane-wave component by summing the remaining terms analytically:
\begin{eqnarray}
    \psi(r)\approx\displaystyle\sum_{l=0}^{l_0}\left[\frac{u_l(r)}{kr}-(2l+1)i^lj_l(kr)\right]P_l(\cos\theta)+e^{ikz}.\nonumber\\
    \label{exactwave}
\end{eqnarray}

Eq.~(\ref{exactwave}) is suitable for low-energy scattering in the absence of Coulomb interactions. 
To derive the wave function including Coulomb effects, we replace the Riccati-Bessel functions $\hat{j}_l$ and $\hat{n}_l$ in Eq.~(\ref{int_of_Schro}) with the Coulomb wave functions $F_l$ and $G_l$~\cite{Math1,Math2}.

By following an analogous procedure to the neutral case, we can obtain the corresponding expressions for $A_l(r)$, $t_l(r)$, and $u_l(r)$ when the Coulomb interaction is taken into account. Finally, the exact wave function in the Coulomb case is given by
\begin{eqnarray}
    &&\psi(r)\approx\displaystyle\sum_{l=0}^{l_0}\left[\frac{u_l(r)}{kr}-e^{i\arg{\Gamma(l+1+i\eta)}}i^l\frac{F_l(kr)}{kr}\right]P_l(\cos\theta)\nonumber\\
    &&+\Gamma(1+i\eta)e^{-\pi\eta/2}e^{ikr\cos{\theta}} \leftindex_1 {F}_1(-i\eta,1,ikr(1-\cos{\theta})),\nonumber\\
    &&
    \label{exactwave_co}
\end{eqnarray}
where ${}_1F_1$ is the confluent hypergeometric function~\cite{Math1,Math2} and $\eta$ denotes the Sommerfeld parameter~\cite{SommerfeldParameter}. The term containing the hypergeometric function represents the Coulomb-distorted plane wave that replaces the free plane wave $e^{ikz}$ in the neutral case.
\subsection{Wave functions in noncentral potentials}
\label{sub:methods_uncen}
Major difficulty of noncentral-force problems faced by the partial wave method lies in coupling between different partial waves. 
For a general noncentral interaction, the possible orbital angular momentum states $L$ that can couple are governed by the angular momentum addition rule~\cite{CGcoeff}
\begin{eqnarray}
\left| {L}-{S} \right| \leq {J} \leq {L}+{S}
\end{eqnarray}
where $J$ is the total angular momentum and $S$ is the total spin. 
In physical applications, the form of the noncentral potential is often constrained by underlying symmetries, which simplifies the coupling structure. 
The most ubiquitous example is the tensor force. This force is associated with a potential term $V_T=S_{12}V_T(r)$, where $S_{12}$ is a rank-2, even-parity, and singlet-annihilating spherical tensor operator~\cite{TensorForce}
\begin{eqnarray}
S_{12}=3\frac{(\vec{\sigma}_1\cdot\vec{r})(\vec{\sigma}_2\cdot\vec{r})}{r^2}-\vec{\sigma}_1\cdot\vec{\sigma}_2.
\label{S12}
\end{eqnarray}
In Eq.~\eqref{S12}, $\vec{\sigma}$ denotes the Pauli matrices, indicating that the current discussion is restricted to spin-1/2 particles. According to the Wigner-Eckart theorem~\cite{WignerEckart}, for a spherical tensor operator of rank $k$, the matrix elements between two coupled partial waves with orbital angular momenta $L$ and $L'$ are subject to the triangular inequality
\begin{eqnarray}
\left| {L}-k \right| \leq {L'} \leq {L}+{k}.
\end{eqnarray}
As an even-parity operator, $S_{12}$ can only couple partial waves that share the same parity $(-1)^L$. Combining these constraints yields the full selection rule for the tensor force operator $S_{12}$
\begin{eqnarray}
\Delta L = 0, 2 ~~~ S=1.
\end{eqnarray}
Assuming the amplitudes of incoming waves in the coupled channels are denoted by $A_1$ and $A_2$ respectively, we can derive the asymptotic form of the radial wave functions in the region where the short-range potentials can be neglected~\cite{Asymp_uw}:
\begin{subequations}
\begin{eqnarray}
&&u(r)=[A_1(1-S_{J,11})-A_2S_{J,12}]\hat{n}_{J-1}\nonumber\\
&&-i[A_1(1+S_{J,11})+A_2S_{J,12}]\hat{j}_{J-1},
\label{int_of_noncenWF1}
\end{eqnarray}
\begin{eqnarray}
&&w(r)=[A_2(1-S_{J,22})-A_1S_{J,12}]\hat{n}_{J+1}\nonumber\\
&&-i[A_2(1+S_{J,22})+A_1S_{J,12}]\hat{j}_{J+1},
\label{int_of_noncenWF2}
\end{eqnarray}
\end{subequations}
where $S_J$ represents the 2$\times$2 scattering matrix in the total angular momentum $J$. In the following, we aim to extend the definitions of $A_1$, $A_2$, and $S_J$ to an arbitrary finite distance.

The coupled radial Schrödinger equations for the case $\Delta L =2$ and $S=1$ take the form~\cite{CoupledRadial}
\begin{subequations}
\begin{eqnarray}
&&\left[ \frac{d^2}{dr^2}-\frac{(J-1)J}{r^2}-U_{J-1,J-1}(r)+k^2 \right] u_{J-1}(r)~~~~~~~~ \nonumber\\
&&=U_{J-1,J+1}(r) u_{J+1}(r),
\label{coupWF1}
\end{eqnarray}
\begin{eqnarray}
&&\left[ \frac{d^2}{dr^2}-\frac{(J+1)(J+2)}{r^2}-U_{J+1,J+1}(r)+k^2 \right] u_{J+1}(r) \nonumber\\
&&=U_{J+1,J-1}(r) u_{J-1}(r),
\label{coupWF2}
\end{eqnarray}
\end{subequations}
where $U_{J\pm 1,J\pm 1}(r)$ denotes the matrix element of the potential operator rescaled by $2\mu$, and is symmetric under interchange of the channel indices. The coupling terms appear on the right hand side of the equations. For brevity, we omit the explicit variable dependence in some functions and adopt the abbreviated notation$$
\begin{array}{cc}
     U_1 \coloneqq U_{J-1,J-1},~U_3\coloneqq U_{J-1,J+1},~U_4\coloneqq U_{J+1,J+1}, \\
     \\
     u\coloneqq u_{J-1},~w\coloneqq u_{J+1}.
\end{array}
$$
Following the formalism developed for the uncoupled case in Eq.~\eqref{int_of_Schro}, we can rewrite the coupled equations in Eqs.~\eqref{coupWF1} and \eqref{coupWF2} in integral form
\begin{subequations}
\begin{eqnarray}
&&u(r)=D\hat{j}_{J-1}(kr)+k^{-1}\int_{0}^{r}ds\left[\hat{j}_{J-1}(kr)\hat{n}_{J-1}(ks)\right.\nonumber\\
&&\left.-\hat{j}_{J-1}(ks)\hat{n}_{J-1}(kr)\right](U_1u(s)+U_3w(s)),
\label{int_of_WF1}
\end{eqnarray}
\begin{eqnarray}
&&w(r)=E\hat{j}_{J+1}(kr)+k^{-1}\int_{0}^{r}ds\left[\hat{j}_{J+1}(kr)\hat{n}_{J+1}(ks)\right.\nonumber\\
&&\left.-\hat{j}_{J+1}(ks)\hat{n}_{J+1}(kr)\right](U_4w(s)+U_3u(s)).
\label{int_of_WF2}
\end{eqnarray}
\end{subequations}
As before, it is convenient to introduce auxiliary functions to simplify the expressions
\begin{subequations}
\begin{eqnarray}
\begin{aligned}
&c\coloneqq D+k^{-1}\int_{0}^{r}dr'\hat{n}_{J-1}(U_1u+U_3w),\\
&s\coloneqq -k^{-1}\int_{0}^{r}dr'\hat{j}_{J-1}(U_1u+U_3w),
\end{aligned}
\label{cs_aux}
\end{eqnarray}
\begin{eqnarray}
\begin{aligned}
&m\coloneqq E+k^{-1}\int_{0}^{r}dr'\hat{n}_{J+1}(U_4w+U_3u),\\
&n\coloneqq -k^{-1}\int_{0}^{r}dr'\hat{j}_{J+1}(U_4w+U_3u),
\end{aligned}
\label{mn_aux}
\end{eqnarray}
\end{subequations}
After performing the substitution and differentiation analogously to Eqs.~\eqref{u_l(r)} and \eqref{deriv_sc}, we proceed to parameterize the $2\times 2$ scattering matrix $S_J$ and extend definition of the associated parameters to finite distances, following the same approach as in Eqs.~\eqref{eqb} and \eqref{eqc}. Two commonly used parameterizations are considered (with the subscript $J$ omitted for brevity)~\cite{Asymp_uw,BarPhaseShift}:
\begin{subequations}
\begin{eqnarray}
\begin{bmatrix}
\cos{\omega}&-\sin{\omega}\\
\sin{\omega}& \cos{\omega}
\end{bmatrix}
\begin{bmatrix}
e^{2i\delta_\alpha}&0     \\
0                  &e^{2i\delta_\beta}
\end{bmatrix}
\begin{bmatrix}
\cos{\omega}&\sin{\omega}\\
-\sin{\omega}&\cos{\omega}
\end{bmatrix}
,
\label{parameter1}
\end{eqnarray}
\begin{eqnarray}
\begin{bmatrix}
e^{2i\eta_\alpha}&0     \\
0                  &e^{2i\eta_\beta}
\end{bmatrix}
\begin{bmatrix}
\cos{2\epsilon}&i\sin{2\epsilon}\\
i\sin{2\epsilon}&\cos{2\epsilon}
\end{bmatrix}
\begin{bmatrix}
e^{2i\eta_\alpha}&0     \\
0                  &e^{2i\eta_\beta}
\end{bmatrix}
,
\label{parameter2}
\end{eqnarray}
\end{subequations}

where $\delta_\alpha$ and $\delta_\beta$ are the eigen phase shifts (also known as the Blatt-Biedenharn phase shifts), $\omega$ is the eigen admixture, $\eta_\alpha$ and $\eta_\beta$ are the bar phase shifts, and $\epsilon$ is the bar admixture.

Through comparison, we find that the parameterization given in Eq.~\eqref{parameter2} leads to significantly simpler forms of the resulting differential equations (see the Appendix for details):
\begin{eqnarray}
&&\eta_\alpha^\prime=\frac{1}{k\cos{2\epsilon}}\left[ U_1\left( -p_0^2\cos^2{\epsilon}+m_0^2\sin^2{\epsilon} \right)\right.\nonumber\\
&&-2U_3\left( -p_0m_2\cos^2{\epsilon}+m_0p_2\sin^2{\epsilon} \right)\nonumber\\
&&\left.+U_4\left( p_2^2\sin^2\epsilon-m_2^2\cos^2\epsilon \right)\right],
\label{D_a}
\end{eqnarray}
\begin{eqnarray}
&&\eta_\beta^\prime=\frac{1}{k\cos{2\epsilon}}\left[ U_4\left( -p_2^2\cos^2{\epsilon}+m_2^2\sin^2{\epsilon} \right)\right.\nonumber\\
&&-2U_3\left( -p_2m_0\cos^2{\epsilon}+m_2p_0\sin^2{\epsilon} \right)\nonumber\\
&&\left.+U_1\left( p_0^2\sin^2{\epsilon}-m_0^2\cos^2{\epsilon} \right)\right],
\label{D_b}
\end{eqnarray}
\begin{eqnarray}
\epsilon^\prime=\frac{1}{k}\left[U_1m_0p_0-U_3\left( p_0p_2+m_0m_2\right)+U_4m_2p_2\right],
\label{MxR}
\end{eqnarray}
where $p_0$, $p_2$, $m_0$, and $m_2$ are real functions defined in Eqs.\eqref{m0p0m2p2}. 

Finally, the exact wave function can be expressed as
\begin{widetext}
\begin{eqnarray}
{}^3\Psi_M=e^{ikz}{}^3\chi_M+\sum^{J_0}_{J=0}\sum^{J+1}_{l=J-1}\left( {}^3u_{l}-\sqrt{4\pi(2l+1)}i^l\frac{\hat{j}_l}{k}\left<l01M|JM\right> \right)\frac{\mathcal{Z}_{l3JM}(\theta,\phi)}{r}
\label{exactwavenoncentral}
\end{eqnarray}
\end{widetext}
which incorporates the Clebsch-Gordan coefficient, the magnetic quantum number $M$ of the total spin, and the total angular momentum wave function $\mathcal{Z}_{lsJM}$~\cite{CoupledRadial}. For coupled channels, the radial wave functions ${}^3u_l$ appearing in Eq.~\eqref{exactwavenoncentral} is related to the variable phase shifts and admixture via $S_J$ in Eq.~\eqref{parameter2}, together with the relation in Eq.~\eqref{int_of_noncenWF1} and \eqref{int_of_noncenWF2}.

One can derive a set of coupled differential equations for the auxiliary functions $A_{1,2}$ introduced above, utilizing Eqs.~\eqref{cs_aux} and \eqref{mn_aux}.

When the Coulomb potential is introduced, the above-mentioned equations remain valid by simply replacing the Riccati–Bessel functions and plane waves with the corresponding Coulomb wave functions, while keeping the rest of the formalism unchanged.

Based on the wave functions derived above, we apply a Lorentz transformation to construct the kernel of the correlation function, $\left| \Psi(\mathbf{q},\mathbf{r^*}) \right|^2$, from which the two-particle correlation functions are then obtained via Eq.~\eqref{CF}.
\section{Results and discussions}
\label{sec:results}
Based on the formulas derived above, we numerically evaluate the two-nucleon correlation functions (CFs) using the Reid soft-core interaction. In these calculations, the actual energy spectra of the emitted nucleons are neglected, and the total momentum $K$ of each pair is accordingly set to zero. The relative momentum is sampled with a uniform bin width of 5~MeV/c.
\subsection{CFs for specific spin and isospin states}
\label{sub:results_A}
The Reid soft-core potentials are defined separately for different total spin $S$ and total isospin $T$ channels.
Accordingly, we classify the corresponding correlation functions by these quantum numbers.
In this paper, the Reid soft-core potentials are specified as follows. For the $T=1$ channels, they include~\cite{ReidPotential}
$$
V({}^1S_0),~ V({}^1D_2),~ V({}^3P_0),~ V({}^3P_1),~ V({}^3P_2-{}^3F_2),
$$
and for the $T=0$ channels, they include
$$
V({}^1P_1),~ V({}^3D_2),~ V({}^3S_1-{}^3D_1).
$$
Here, the letters $S,P,D,$ and $F$ denote the orbital angular momentum quantum numbers. The presuperscript indicates the spin multiplicity ($2S+1$), while the subscript denotes the total angular momentum $J$.

Moreover, as indicated by Eq.~\eqref{exactwavenoncentral}, the magnetic quantum number $M$ also influences the asymptotic boundary conditions and thus affects the resulting correlation functions. We therefore adopt $M$ as an additional classification parameter. The overall correlation function is obtained by a weighted average over all relevant quantum number configurations. For unpolarized emission, the weighting factors are identical.
\begin{figure}
\centering
\includegraphics[width=0.5\textwidth]{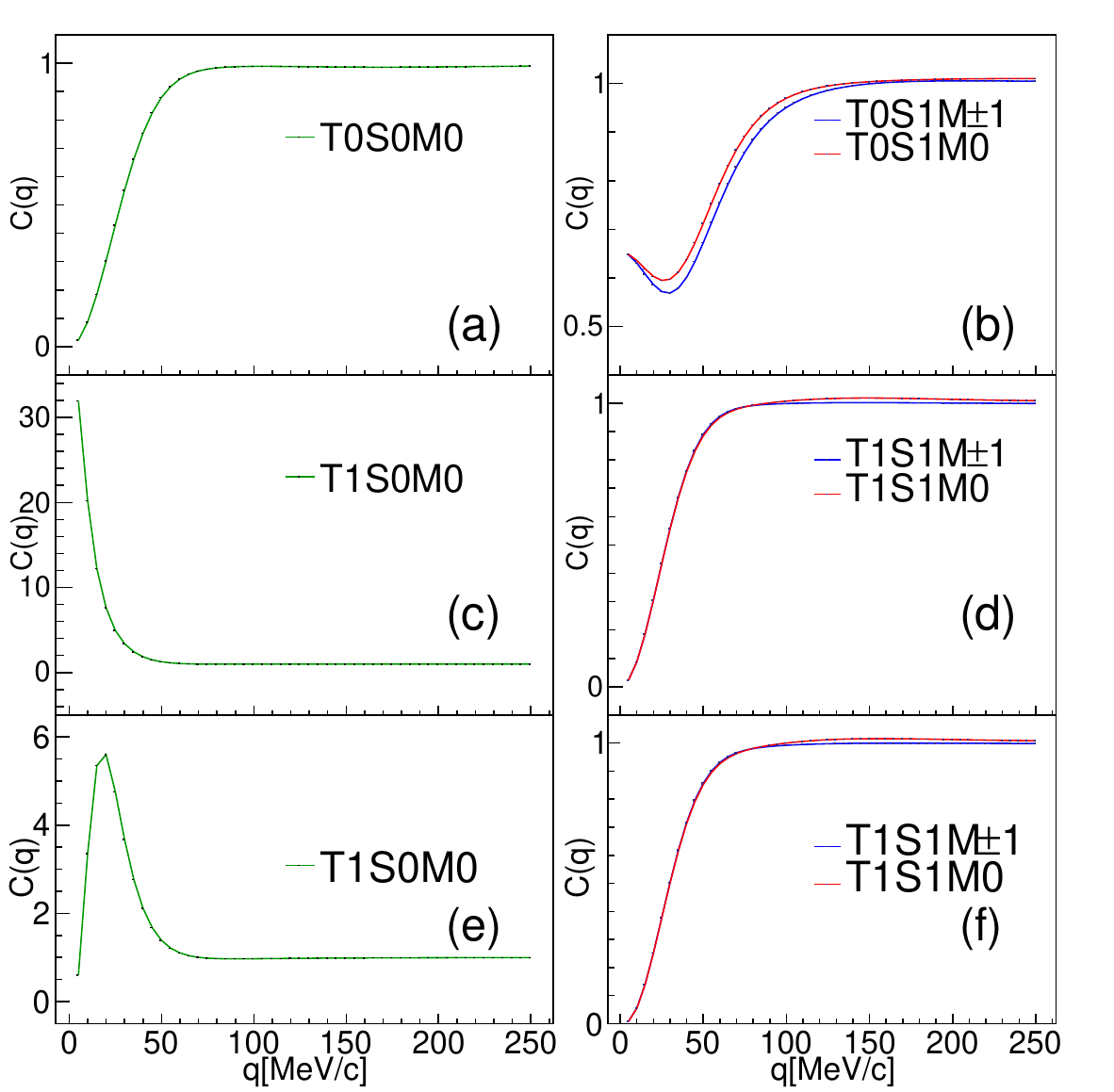}
\caption{
Classified correlation functions for a Gaussian source of 3~fm, with specified isospin $T$, spin $S$, and magnetic quantum number $M$.
(a–d) Neutron–proton correlation functions in the $T=0,S=0$ (a), $T=0,S=1$ (b), $T=1,S=0$ (c), and $T=1,S=1$ (d) channels.
(e,f) Proton–proton correlation functions in the $T=1,S=0$ (e) and $T=1,S=1$ (f) channels. 
}
\label{fig:pads_np_pp}
\end{figure}
Fig.~\ref{fig:pads_np_pp} (a-d) present the classified correlation functions for neutron-proton (n-p) pairs with a Gaussian source of 3~fm. The results exhibit considerable variation across different combinations of $S$ and $T$, primarily due to the interplay of quantum statistics and the structure of the nuclear interaction. In contrast, the dependence on $M$ is relatively weak, leading to only minor deviations among correlation functions with different $|M|$ values. The magnitude of these deviations is also correlated with $T$, and is particularly more pronounced for the $T=0$ channels due to the presence of the low-lying coupled channel ${}^3S_1-{}^3D_1$. The influence of each channel on the correlation functions will be discussed in the next subsection.

Fig.~\ref{fig:pads_np_pp} (e) and Fig.~\ref{fig:pads_np_pp}(f) present the classified correlation functions for proton-proton (p-p) pairs. The influence of the Coulomb force is more prominent in the $S=0$ channel, whereas in the $S=1$ channel its effects are masked by quantum statistics. The neutron-neutron (n-n) correlation functions are not shown, because they are identical to the $T=1$ channel of the n-p correlation functions under the isospin symmetry.

The considerable differences among the classified correlation functions call for increased prudence when selecting the weighting factors. The rationality of using identical weighting factors for nucleons is ensured by thermal equilibrium of the system and multiple collisions in the hadron gas~\cite{ThermalEqui_Polar1,ThermalEqui_Polar2}, which are two sufficient conditions applicable to most circumstances. Otherwise~\cite{ThermalEqui_Polar3}, the weighting factors for different total spin channels should be treated as independent variables or additional fitting parameters.
\subsection{Contributions of each partial waves}
To evaluate the impact of higher partial waves, we construct the correlation function by adding the Reid potentials of a given isospin channel to free particle system in an increasing order of total angular momentum $J$. With an assumption of unpolarized emission, the resulting modification $\Delta$CF to the averaged correlation function under a specific $T$ is examined.
\begin{figure*}
    \centering
    \includegraphics[width=0.9\textwidth]{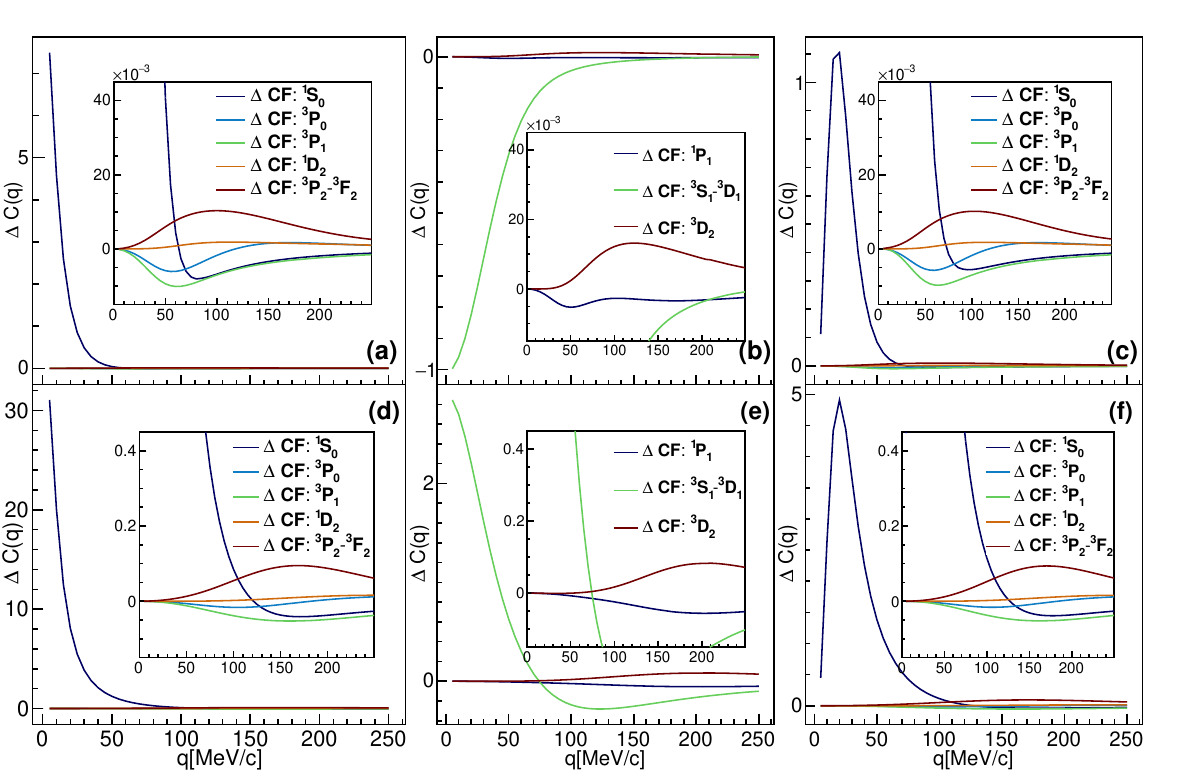}
    \caption{Variations of the correlation functions upon sequentially adding Reid potentials of specific channels. 
             (a,b) Neutron–neutron correlation functions in the $T=1$ (a) and $T=0$ (b) channels with a 3~fm Gaussian source. 
             (c) Proton–proton correlation function with a 3~fm Gaussian source. 
             (d,e) Neutron–neutron correlation functions with a 1~fm Gaussian source. 
             (f) Proton–proton correlation function with a 1~fm Gaussian source. 
             The addition order follows the legend from top to bottom. Zoomed regions are shown to better display the contributions from higher partial waves.}
    \label{fig:Diff}
\end{figure*}

By summarizing the features in Fig.\ref{fig:Diff} (a-c), we find that partial waves with the lowest orbital angular momentum dominate the correlation function at low relative momenta. In contrast, the contributions from higher partial waves peak at relatively large momenta and gradually diminish as the relative momentum decreases. This behavior is consistent with the general property that partial scattering amplitudes decrease rapidly with increasing angular momentum in most low-energy scattering scenarios. However, as shown in Fig.~\ref{fig:Diff}(b), the contribution from ${}^3D_2$ is larger than that from ${}^1P_1$. This exception arises from the higher statistical weight of the triplet state under the assumption of unpolarized emission.

For a Gaussian source of 3~fm, the magnitude of these higher partial wave contributions is around or below 0.01, rendering them indistinguishable in most experimental measurements. Comparing Fig.~\ref{fig:Diff} (a) and Fig.\ref{fig:Diff} (c), the introduction of the Coulomb potential only affects the behavior of $\Delta$CF at very low relative momenta, leaving the contributions from higher partial waves essentially unchanged. 

To enhance the visibility of higher partial wave contributions, we reduce the source size to a smaller value. As illustrated in Fig.~\ref{fig:Diff} (d-f), the tighter spatial extent increases sensitivity to short-range components of the interaction, thereby amplifying the influence of higher angular momentum channels by almost an order of magnitude. Correspondingly, the peaks of their contributions shift toward higher relative momenta, as can be inferred from Eq.~\ref{CF}. For a Gaussian source and zero total momentum $K$, the peak of a given partial wave contribution occurs when the peak of the radial source distribution $r^2\exp{[-r^2/(2\sigma)^2]}$ maximally overlaps with the absolute square of the radial wave function $|u_l(r)|^2$. As the source size decreases, the incident momentum $q$ increases, causing the peaks of $|u_l(r)|^2$ to shift leftward along with the radial source distribution. For noncentral forces, the coupled radial wave functions behave similarly. 

Unlike other curves in Fig.~\ref{fig:Diff} that exhibit monotonic behavior, a flip between positive and negative correlation is observed in Fig.~\ref{fig:Diff} (e) and Fig.~\ref{fig:Diff} (b) as the Gaussian source size increases from 1~fm to 3~fm. As the source further increases from 4~fm to 5~fm, the negative correlation weakens instead. These features indicate a well‑known characteristic of the ${}^3S_1-{}^3D_1$ channel, namely near‑threshold resonance scattering, which arises from a shallow bound state known as the deuteron. In the low-energy limit $k\to 0$ and in the intermediate region, the radial wave function of the ${}^3S_1$ channel is dominated by the scattering length $a$, such that $u(r)\approx k(r-a)$~\cite{Linearu_0}.
Consequently, the kernel of the correlation function under these conditions is given by
$$
\left| \Psi(\mathbf{0},\mathbf{r}) \right|^2\approx2\left(2-\frac{a}{r}\right)^2.
$$

\begin{figure} [hbt]
\centering
\includegraphics[width=0.4\textwidth]{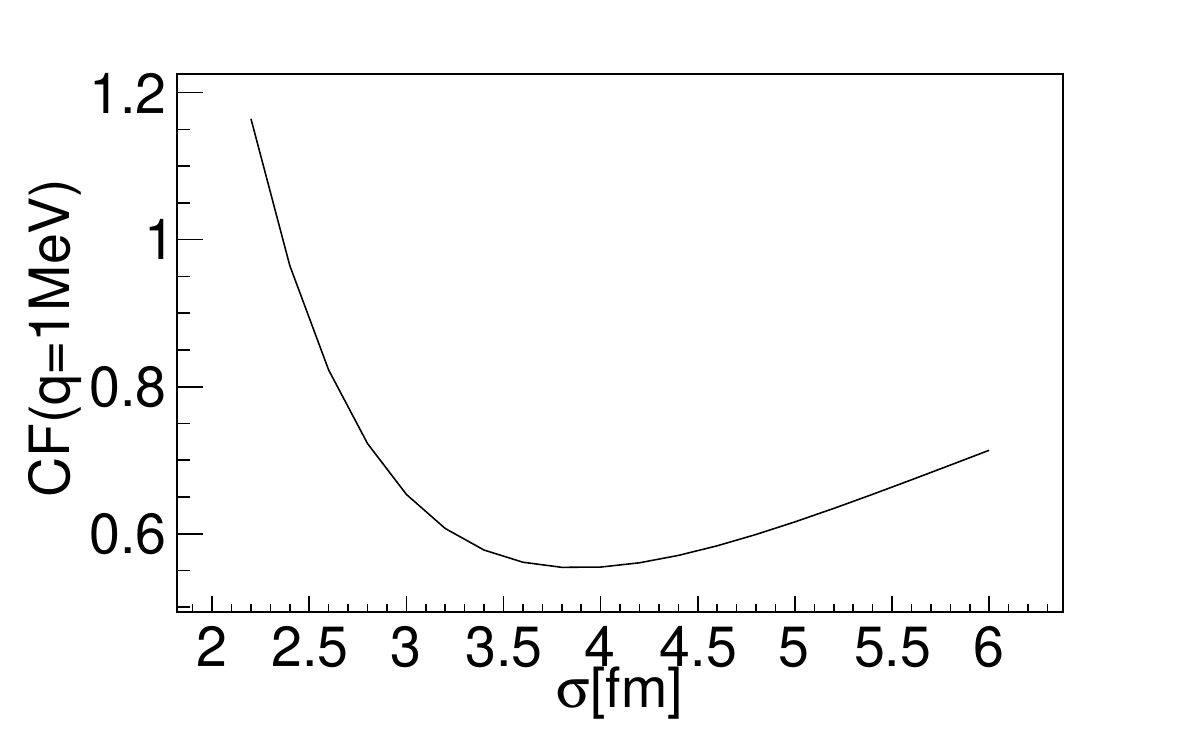}
\caption{
The neutron-proton correlation function for the $T=0$, $S=1$ channel at a relative momentum of $q=1$~MeV as a function of the Gaussian source size $\sigma$.
}
\label{fig:ResonanSS}
\end{figure}

\begin{figure*}[hbt]
\centering
\includegraphics[width=0.9\textwidth]{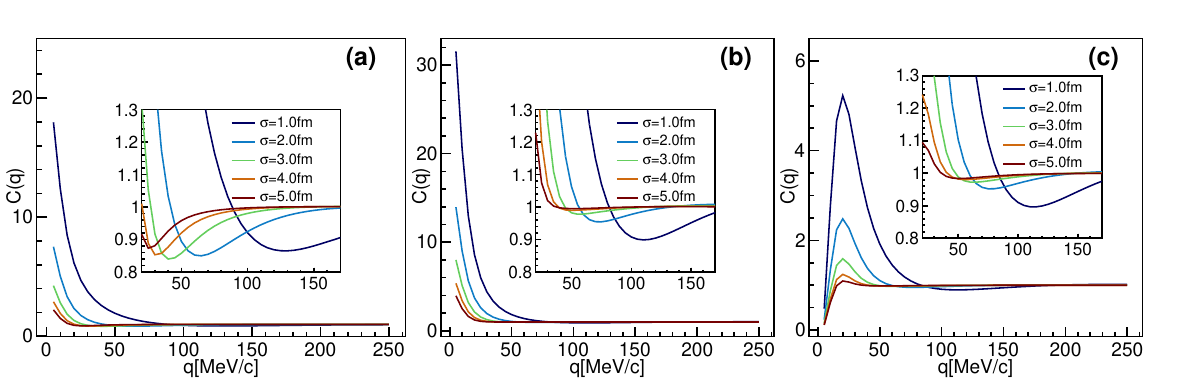}
\caption{
Correlation functions for Gaussian sources with different standard deviation $\sigma$.
(a) Neutron-proton correlation functions. 
(b) Neutron-neutron correlation functions. 
(c) Proton-proton correlation functions. 
}
\label{fig:SS}
\end{figure*}

The presence of resonance leads to a large $a$ (5.39~fm) and an extended intermediate region where the radial wave function remains approximately linear. With this approximation, the correlation function reaches a minimum when the overlap between the source distribution and the kernel is minimal, and then rebounds as the source size continues to increase, as shown in Fig.~\ref{fig:ResonanSS}. In contrast, for other ordinary non-resonant scattering channels such as ${}^1S_0$, the scattering length $a$ is negative ($-7.78$~fm), which leads to a monotonic decrease of correlation functions as the source size increases.
\subsection{Influence from Gaussian source size}
\label{sub:results_C}
The size of the emission source is a critical parameter that depends on the collision system and the pair type being analyzed. In most circumstances, its value can range from below 1~fm in $\mathrm{e^+}-\mathrm{e^-}$ annihilations~\cite{SmallSource} to tens of femtometers in low-energy nuclear collisions~\cite{BigSource}. To avoid the significant time-delay effect occurring in low-energy collisions~\cite{TimeDelay}, we focus on a size range of 1$\sim$5~fm, which is representative of the typical scales encountered in p-p and high-energy heavy-ion collisions~\cite{DependenceofSourceSize}.

Fig.~\ref{fig:SS} presents the overall correlation functions under the assumption of unpolarized emission. As noted earlier, a smaller source implies stronger nuclear interaction during the early stage of emission. Mathematically, this can also be understood as a narrower sampling range of the wave function. Both perspectives account for the more drastic variations observed in the curves of Fig.~\ref{fig:SS} as the source size decreases.

Another noteworthy feature is the recoil structure of the correlation functions, shown in the zoomed regions of Fig.~\ref{fig:SS}.
It is evident from Fig.~\ref{fig:Diff} that all partial waves contribute to the shape of this recoil. For n-n (panel b) and p-p (panel c) pairs, the recoil depth increases monotonically as the source size decreases, consistent with the trend observed in the peak heights. In contrast, the n-p correlation function (panel a) exhibits a different behavior, with the recoil depth reaching a maximum at around $\sigma = 3$~fm. This difference arises from the distinctive nature of the ${}^3S_1-{}^3D_1$ channel, as discussed in the previous subsection.

\section{Conclusions}
\label{sec:discussion}
In this article, we extend the variable phase method to the case of noncentral forces for spin-1/2 particles. Within this framework, the phase shifts and the admixture can be obtained without solving the original Schrödinger equation. Furthermore, we develop a new approach to derive the exact wave function by reducing the second order ordinary differential equation to two first order ordinary differential equations with mutually independent boundary conditions, thereby avoiding the trial cost inherent in the shooting method~\cite{ShootingMethods1,ShootingMethods2}.

Based on this framework, we conduct an in depth investigation of nucleon-nucleon correlation functions. The main topics include: (1) the behavior of correlation functions under specific spin and isospin channels, (2) the contribution of each partial wave to the correlation functions, and (3) the influence of the Gaussian source size on the correlation functions. From these investigations, we summarize several of the most interesting conclusions:
\begin{enumerate}
\item The differences among correlation functions in specific channels highlight the importance of their weighting factors.
\item To make the contributions of higher order partial waves to the correlation function exceed 0.1, the source size should be around or below 1~fm.
\item The near threshold resonance scattering in the ${}^3S_1-{}^3D_1$ channel has a significant effect on the recoil depth of the neutron-proton correlation function and relatively reduces its dependence on the source size.
\end{enumerate}

Our framework can be readily extended to other two-particle correlations (e.g., $\Lambda\Lambda$, p$\Lambda$) with various potentials, as well as to other application scenarios that require the derivation of phase shifts or exact wave functions. 

\textbf{{Acknowledgement}} This work is supported by the National Natural Science Foundation of China under Grant Nos. 12335008 and 11927901, by the Ministry of Science and Technology under Grant No. 2022YFE0103400,  by the Center for High Performance Computing and Initiative Scientific Research Program in Tsinghua University.

\onecolumngrid
\appendix
\section{Derivation of the extended variable phase equations in the noncentral force}\label{app:proof}
In this section, the derivation process of Eq.~\eqref{D_a}, \eqref{D_b} and \eqref{MxR} is detailedly demonstrated.

To associated the auxiliary functions $c$, $s$, $m$, and $n$ with the scattering parameters in Eqs.~\eqref{parameter1} and \eqref{parameter2}, we substitute the Eqs.~\eqref{cs_aux} and \eqref{mn_aux} into the coupled radial wave functions Eqs.~\eqref{int_of_noncenWF1} and \eqref{int_of_noncenWF1}. Through comparison, we find the correspondence
\begin{subequations}
\begin{eqnarray}
\begin{aligned}
&s=A_1-S_{J,11}A_1-S_{J,12}A_2,\quad ic=A_1+S_{J,11}A_1+S_{J,12}A_2,
\end{aligned}
\label{s and ic}
\end{eqnarray}
\begin{eqnarray}
\begin{aligned}
&n=A_2-S_{J,22}A_2-S_{J,21}A_1,\quad im=A_2+S_{J,22}A_2+S_{J,21}A_1.
\end{aligned}
\label{n and im}
\end{eqnarray}
\end{subequations}
We represent $A_1$ and $A_2$ with the auxiliary functions via Eqs.~\eqref{s and ic} and ~\eqref{n and im}, and introduce $B_1$ and $B_2$ to match the degree of freedom:
\begin{subequations}
\begin{eqnarray}
\begin{aligned}
&A_1=(s+ic)/2,\quad A_2=(n+im)/2,
\end{aligned}
\label{A_1 and A_2}
\end{eqnarray}
\begin{eqnarray}
\begin{aligned}
&B_1=(ic-s)/2,\quad B_2=(im-n)/2.
\end{aligned}
\label{B_1 and B_2}
\end{eqnarray}
\end{subequations}
Then we use $\alpha$, $\beta$ and $\gamma$ to denote the components of $S_J$ and express them in the form of $A_1$, $A_2$, $B_1$ and $B_2$:
\begin{subequations}
\begin{eqnarray}
\begin{aligned}
\alpha\coloneqq S_{J,11}=\frac{B_1B^\star_2-A_1^\star A_2}{A_1B^\star_2-B_1^\star A_2},
\end{aligned}
\label{alpha}
\end{eqnarray}
\begin{eqnarray}
\begin{aligned}
\beta\coloneqq S_{J,22}=\frac{A_1A^\star_2-B_1^\star B_2}{A_1B^\star_2-B_1^\star A_2},
\end{aligned}
\label{beta}
\end{eqnarray}
\begin{eqnarray}
\begin{aligned}
\gamma\coloneqq S_{J,12}=-\frac{\left|A_2\right|^2-\left|B_2\right|^2}{A_1B^\star_2-B_1^\star A_2}=-\frac{\left|A_1\right|^2-\left|B_1\right|^2}{A_1B^\star_2-B_1^\star A_2}.
\end{aligned}
\label{gamma}
\end{eqnarray}
\end{subequations}
A possible overdetermination can be discovered in Eq.~\eqref{gamma}. So for the self-consistency of $\gamma$, these two terms should be equal, which is equivalent to the proof of $mn^\star+cs^\star-nm^\star-sc^\star=0$. As $r\to 0$, this expression tends to zero according to Eqs.~\eqref{mn_aux} and \eqref{cs_aux}. Then we only need to prove that its derivative remains zero for arbitrary $r$:
\begin{eqnarray}
\begin{aligned}
(mn^\star+cs^\star-nm^\star-sc^\star)^\prime&= -k^{-1}({n}_2n+{j}_2m)[U_3({n}_0s^\star+{j}_0c^\star)+U_4({n}_2n^\star+{j}_2m^\star)]\nonumber\\
&
~~~~+k^{-1}({n}_2n^\star+{j}_2m^\star)[U_3({n}_0s+{j}_0c)+U_4({n}_2n+{j}_2m)]\nonumber\\
&
~~~~-k^{-1}({n}_0s+{j}_0c)[U_1({n}_0s^\star+{j}_0c^\star)+U_3({n}_2n^\star+{j}_2m^\star)]\nonumber\\
&
~~~~+k^{-1}({n}_0s^\star+{j}_0c^\star)[U_1({n}_0s+{j}_0c)+U_3({n}_2n+{j}_2m)]\nonumber\\
&
=0\nonumber,
\end{aligned}
\end{eqnarray}
where the derivative of the auxiliary functions is the noncentral version of Eq.~\eqref{deriv_sc} derived by Eqs.~\eqref{cs_aux} and \eqref{mn_aux}, and ${j}_0$, ${j}_2$, ${n}_0$, and ${n}_2$ are abbreviations of $\hat{j}_{J-1}$, $\hat{j}_{J+1}$, $\hat{n}_{J-1}$, and $\hat{n}_{J+1}$ respectively.

Next, we denote the phase shifts and admixtures in terms of $\alpha$, $\beta$, and $\gamma$. For the two different sets of parameters in Eqs.~\eqref{parameter1} and \eqref{parameter2}, the expressions are written as
\begin{subequations}
\begin{eqnarray}
\left\{
\begin{aligned}
e^{2i\delta_{\alpha,\beta}}&=\frac{\alpha+\beta\pm\sqrt{4\gamma^2+(\alpha-\beta)^2}}{2}\\
\tan{2\omega}&=\frac{2\gamma}{\alpha-\beta}
\end{aligned}
\right.,
\label{eigen denote by xilazimu}
\end{eqnarray}
\begin{eqnarray}
\left\{
\begin{aligned}
&e^{4i\eta_\alpha}=\alpha/\alpha^\star, \quad e^{4i\eta_\beta}=\beta/\beta^\star\\
&\tan^2{2\epsilon}=-\frac{\gamma^2}{\alpha\beta}
\end{aligned}
\right. .
\label{bar denote by xilazimu}
\end{eqnarray}
\end{subequations}
In comparison of these two sets, it is obvious that Eq.~\eqref{bar denote by xilazimu} is much easier for differentiation. So we take the derivative of both sides of Eq.~\eqref{bar denote by xilazimu}, and express the right side by the eigen phase shifts, eigen admixture, and known functions. To conveniently realize the purpose of expressing, we introduce the second-level auxiliary functions $\Phi$, $\Delta$, and $\phi$:
\begin{eqnarray}
\Phi\coloneqq B_1B_2^\star-A_1^\star A_2,\quad\Delta\coloneqq A_1B_2^\star-B_1^\star A_2,\quad\phi\coloneqq-\left|A_2\right|^2+\left|B_2\right|^2,
\label{second-level aux}
\end{eqnarray}
through which the components of $S_J$ can be written as $\alpha=-\Phi^\star/\Delta$, $\beta=\Phi/\Delta$, and $\gamma=\phi/\Delta$, and the derivatives of $\alpha$ ,$\beta$, and $\gamma$ are transformed into that of the second-level auxiliary functions:
\begin{subequations}
\begin{eqnarray}
\begin{aligned}
k\Phi^\prime=&U_1\left[ in_0^2\left( \Delta^\star+\Phi \right)+ij_0^2\left( \Phi-\Delta^\star \right)-2j_0n_0\Delta^\star\right]
            +2U_3\phi\left( -in_0n_2+j_0n_2-j_2n_0-ij_0j_2 \right)\\
            +&U_4\left[ in_2^2\left( \Delta-\Phi \right)-ij_2^2\left( \Phi+\Delta \right)+2j_2n_2\Delta \right],
\end{aligned}
\end{eqnarray}
\label{dPhidx}
\begin{eqnarray}
\begin{aligned}
k\Delta^\prime=&U_1\left[ in_0^2\left( \Phi^\star+\Delta \right)+ij_0^2\left( \Delta-\Phi^\star \right)-2j_0n_0\Phi^\star\right]
            +2U_3\phi\left( -in_0n_2+j_0n_2+j_2n_0+ij_0j_2 \right)\\
            +&U_4\left[ in_2^2\left( \Delta-\Phi \right)+ij_2^2\left( \Phi-\Delta \right)+2j_2n_2\Phi \right],
\end{aligned}
\end{eqnarray}
\label{dDeltadx}
\begin{eqnarray}
\begin{aligned}
2k\phi^\prime=U_3&\left[in_0n_2\left( -\Phi^\star+\Delta^\star+\Phi-\Delta \right)+ij_0j_2\left( -\Phi^\star-\Delta^\star+\Phi+\Delta \right)\right. \\
&\left.+j_0n_2\left( \Phi^\star-\Delta^\star+\Phi-\Delta \right)+j_2n_0\left( -\Phi^\star-\Delta^\star-\Phi-\Delta \right)\right].
\end{aligned}
\end{eqnarray}
\label{dgammadx}
\end{subequations}
The second-level auxiliary functions in the right side of the derivatives of Eqs.~\eqref{bar denote by xilazimu} can be replaced and fully canceled, thus leading to the final noncentral variable phase equations. Take the differential equation regarding $\epsilon$ as an example:
\begin{eqnarray}
\begin{aligned}
\epsilon^\prime&= \frac{1}{4k}\left\{ U_1\left[ -n_0^2\sin(2\eta_\alpha)\sin(2\epsilon)+j_0^2\sin(2\eta_\alpha)\sin(2\epsilon)-2j_0n_0\cos(2\eta_\alpha)\sin(2\epsilon) \right] \right.\\
&\qquad+2U_3
\begin{aligned}[t]
&\left[ -j_0j_2\left( \cos(2\epsilon)\cos(\eta_\alpha+\eta_\beta)+\cos(\eta_\alpha-\eta_\beta) \right)\right.
                +n_0n_2\left( \cos(2\epsilon)\cos(\eta_\alpha+\eta_\beta)-\cos(\eta_\alpha-\eta_\beta) \right) \\
                &+j_0n_2\left( \sin(\eta_\alpha-\eta_\beta)-\cos(2\epsilon)\sin(\eta_\alpha+\eta_\beta) \right)
         \left. -j_2n_0\left( \cos(2\epsilon)\sin(\eta_\alpha+\eta_\beta)+\sin(\eta_\alpha-\eta_\beta) \right) \right]  \\
\end{aligned}
&\\
&
\qquad+\left.U_4\left[ -n_2^2\sin(2\eta_\beta)\sin(2\epsilon)+j_2^2\sin(2\eta_\beta)\sin(2\epsilon)-2j_2n_2\cos(2\eta_\beta)\sin(2\epsilon) \right] \right\}.
\end{aligned}
\label{origin epsilon}
\end{eqnarray}
Severe floating-point precision issues emerge as $kr\to 0$ and the second kind of Riccati-Bessel functions tend to infinity. To solve this problem, we simplify the trigonometric functions in Eq.~\eqref{origin epsilon} by introducing $m_0$, $p_0$, $m_2$, and $p_2$:
\begin{eqnarray}
\begin{aligned}
&m_0=(j_0\sin{\eta_\alpha}-n_0\cos{\eta_\alpha})\sin{\epsilon},\quad p_0=(j_0\cos{\eta_\alpha}+n_0\sin{\eta_\alpha})\cos{\epsilon},\\
&m_2=(j_2\sin{\eta_\beta}-n_2\cos{\eta_\beta})\sin{\epsilon},\quad p_2=(j_2\cos{\eta_\beta}+n_2\sin{\eta_\beta})\cos{\epsilon},\\
\end{aligned}
\label{m0p0m2p2}
\end{eqnarray}
through which we are able to derive the Eqs.~\eqref{D_a}, \eqref{D_b}, and \eqref{MxR}.

\twocolumngrid
\bibliographystyle{apsrev4-2}  
\bibliography{apssamp}

\end{document}